# PROBABILISTIC SCHMID FACTORS AND SCATTER OF LCF LIFE

H. Gottschalk[1], S. Schmitz[2], T. Seibel[3], G. Rollmann[2], R. Krause[4] and T. Beck[5]

[1] Bergische Universität Wuppertal (FBC), Postfach 42097, 42119 Wuppertal, Germany
[2] Siemens AG, Mellinghoferstr 55, 45473 Mülheim, Germany
[3] Forschungszentrum Jülich, Wilhelm-Johnen-Straße, 52425 Jülich, Germany
[4] Universita della Svizzera Italiana, Via Guiseppe Buffi 13, 6904 Lugano, Switzerland
[5] TU Kaiserslautern, Gottlieb-Daimler-Straße, 47663 Kaiserslautern, Germany

**ABSTRACT**

We investigate the scatter in the cycles to crack initiation for conventionally cast specimens of the superalloy RENE 80 in total strain controlled low cycle fatigue (LCF) tests at 850°C. The grains at the location of crack initiation are investigated using electron backscatter diffraction (EBSD) measurements with scanning electron microscopy (SEM). This results in determination of maximal Schmid factors for the slip system that lead to the initiation of the LCF surface crack. It is shown using stress – life and strain – life models, that taking into account the Schmid factor considerable reduces the scatter in LCF life prediction. This is used to propose a probabilistic model for LCF life based on random grain orientations. Using Mote Carlo simulation, we determine the probability distribution for the maximal Schmid factor in an uniaxial stress state under isotropic random orientations of the grain. This is used to calculate failure probabilities under the above mentioned model and to compare them with experimental data. Based on extreme value statistics, it is discussed whether a high maximal Schmid factor can be considered as causally related to an early initiation of an LCF crack and we present some rather surprising experimental evidence. In the last section conclusions are formulated and future directions of research are outlined.

**INTRODUCTION**

Even under lab conditions, the number of load cycles to the initiation of fatigue cracks for fixed load amplitudes underlies a statistical scatter up to a factor 10. This is in particular true for coarse grained cast Ni – based superalloys, where forging is not an option to generate fine, random grain structures which could reduce scatter in crack initiation life. This lack of predictiveness has severe economical consequences in the operation of gas turbines, both in aviation and energy production. Any quantitative description of the crack initiation probabilities for gas turbines would therefore greatly benefit airlines and power plant operators as it would provide input for risk based servicing activities.

The statistical investigation of fatigue has been studied in a series of papers, see e.g. [1-3]. In a previous article we have proposed a probabilistic model for LCF crack initiation based on the usual Ramberg-Osgood and Coffin-Manson-Basquin equaltions in combination with local Weibull analysis [4]. While this model has been calibrated with strain amplitude – fatigue life data and has been applied to the design of stationary gas turbines [5-7], it follows an empirical and statistical approach and should be further elaborated from a material science prospective.

It is therefore desirable to investigate further the physical origin of the enormous scatter in LCF life. As the intra granular $\gamma'$-precipitate / $\gamma$-matrix structure is rather regular, the orientation of the face centred cubic $\gamma$-phase relative to the stress state $\sigma$ seems to be an interesting candidate. This concept leads to the calculation of the highest shear stress acting on one of the twelve slip systems of the grain, i.e. the Schmid factor times the nominal stress [7-14].

Working with this hypothesis, in a first step we need to calculate the probability distribution of Schmid factors under isotropic grain orientations. Here we do this with a Monte Carlo approach. In a second step, the effect of Schmid factor orientations to the specimen LCF life needs to be investigated. Scatter plots of life vs stress and strain, respectively, show a considerable reduction in statistical scatter if the stress replaced by stress corrected by the Schmid factor [7]. Combining the probability distribution of Schmid factors with a simple model of strain correction based on the Coffin-Manson Basquin equation, we find that the scatter originating from the variation of the Schmid factor of a single grain has about the same order of magnitude as the experimentally observed scatter.

There are however two observable effects that do irritate in the seemingly consistent picture described above: First, if the Schmid factor alone would be responsible for the selection of a grain at the specimen surface that initiates the crack, the relevant distribution of Schmid factors would not be the isotropic Schmid factor distribution, but it's related extremal value distribution of the order of grains on the specimen surface. This distribution however is much more peaked at high Schmid factors and would largely underestimate the resulting scatter in LCF-life. Even more irritating, the experimental Schmid factors measured by electron backscatter diffraction (EBSD) at those grains where LCF-cracks actually initiated do not show any tendency towards high values compared with the isotropic Schmid factor distribution. This prohibits seeing high Schmid factors in simple a causal relation to crack initiation. Possible consequences from the experimental findings for future modelling of scatter in LCF-life are being discussed.

# SPECIMEN, MATERIAL, TESTING AND FRACTOGRAPHY

**Material**

Total strain controlled LCF tests have been conducted at 850°C with specimens from the Ni-base alloy RENE 80, confer Table 1. The material has been provided by Siemens Energy in form of conventionally cast plates with dimensions 200 x 112 x 20mm. Three different specimen geometries have been used in order to account for statistical size effects, see Figure 1. The geometrical data of the gauge length are summarized in Table 2.

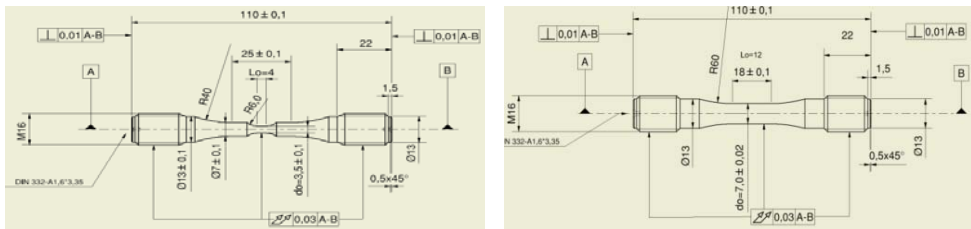

Fig. 1: Two out of three specimen geometries

Table 1: Chemical composition of RENE80 (in weight %)

| Element | Ni | Cr | Co | Ti | Mo | W | Al | C | B | Zr |
|---|---|---|---|---|---|---|---|---|---|---|
| Weight-% | 60 | 14 | 9.5 | 5 | 4 | 4 | 3 | 0.7 | 0.015 | 0.03 |

Table 2: Specimen geomety data (gauge area)

| Geometry | $L_0$ [mm] | $D_0$ [mm] | Surface [mm$^2$] |
|---|---|---|---|
| G1 | 18 | 7 | 395,84 |
| G2 | 4 | 3.5 | 43,98 |
| G3 | 18 | 11 | 622.04 |

**Testing**

LCF tests were conducted on a servohydraulic test rig with maximal load capacity of 100kN. 14 Tests have been performed with geometry G1, 13 with G2 and eight with G3 at strain amplitude levels between 0.1% and 0.3% at load ratio $R_\epsilon = -1$. The load cycle shape was chosen triangular at $0.1 - 1Hz$ as depending on the load amplitude. The specimen size was taken into account when determining the amount of stress drop that indicates crack initiation [7].

**Fractography**

The inspection of fracture surfaces with SEM show that all cracks initiated at the surface of the specimen. Cracks initiated at slip planes in surface grains as it is often observed in coarse grained materials, and in Ni-based superalloys with $\gamma/\gamma'$ - microstructure, in particular, cf. Figure 2 [4].

At low and medium strain amplitudes, a relatively smooth crack near the crack initiation point is followed by a zone with stable crack growth marked by semicircular crack fronts. The final crack causing spontaneous failure shows a rough surface morphology. In contrast to this, in the case of high strain amplitudes, multiple crack initiation points and global plastification in the specimen can be observed.

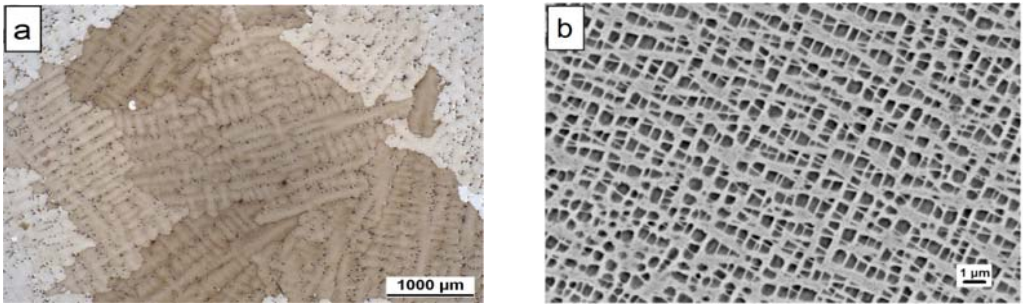

Fig. 2: Grain structure (OM) and internal $\gamma/\gamma'$-structure (SEM) in RENE 80

Lattice orientations in the grain at the main crack initiation point have been determined via backscatter electron diffraction with a SEM for all 40 specimens. Based on the orientation, the Schmid factor for the slip system associated to the initial crack plane has been determined.

**SCHMID FACTORS AND LCF LIFE**

The $\gamma$-phase matrix (Figure 2b) in the Ni-based alloy RENE 80 exposes a face centred cubic (fcc) lattice structure (Figure 4a). The tetrahedral fcc symmetry leads to four slip planes identical to the surfaces of the tetrahedra and slip directions parallel to their three edges (Figure 4b).

The shear stress within the slip plane with normal $n_i, i = 1,2,3,4$, in the direction of the slip system $s_{i,j}, j = 1,2,3$, given the stress tensor $\sigma$ is

$$\tau_{i,j} = n_i \cdot \sigma \cdot s_{i,j}, \tau = max|\tau_{ij}| \qquad (1)$$

with $\tau$ being the maximal shear stress on one of the 12 slip systems. In specimen testing the stress state $\sigma$ is homogeneous and uniaxial in n-direction where $e$ is a vector of unit length

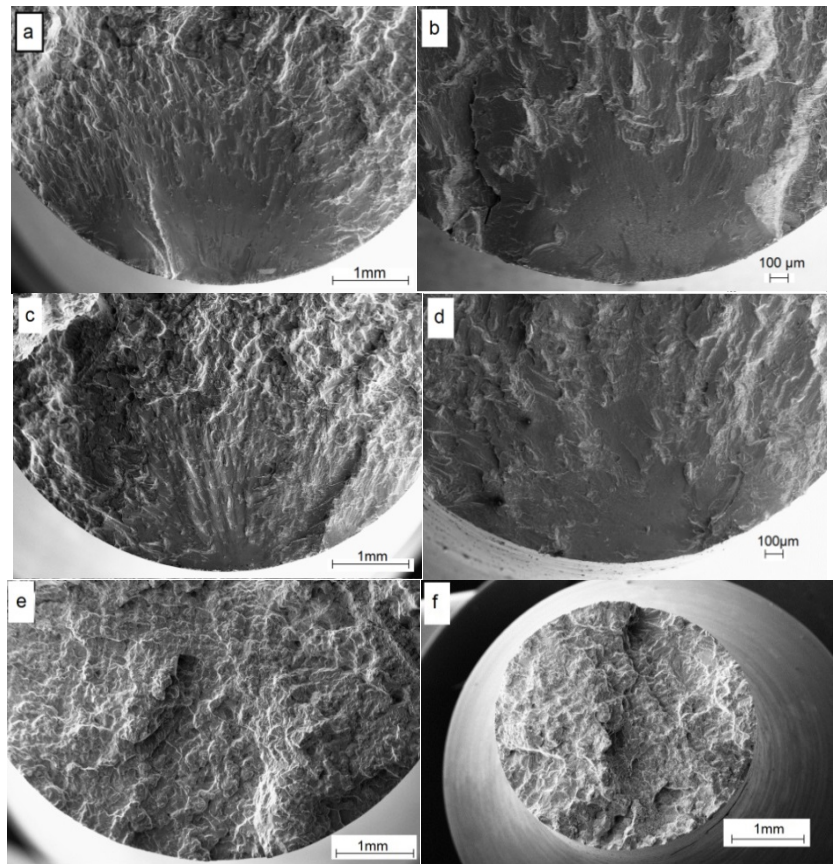

Fig. 3: Typical crack planes for standard specimen G1 (a,c,e) and small specimen G2 (b,d,f). a-b have small, c-d medium and e-f high load amplitude.

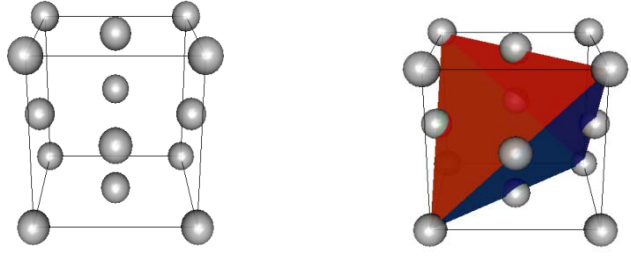

Fig. 4: FCC lattice (left) with four slip planes and twelve slip systems at the edges of the triangle (right)

$$\sigma = \sigma_I e \otimes e, \quad \text{with } (e \otimes e)_{i,j} = e_i e_j \tag{2}$$

The maximal Schmid factor in this case is defined as $m = \tau/\sigma_I$.

The relative orientation between the grain and the direction of the uniaxial stress, $n = e_z$, the unit vector in z-direction is given by a 3x3-orthogonal matrix $U$ that transforms slip plane and slip system orientation:

$$n_i \rightarrow U n_i, s_{i,j} \rightarrow U s_{i,j} \tag{3}$$

Therefore, we can calculate U-dependent Schmid factors for the slip system $n_i$ and $s_{ij}$,

$$\tau_{i,j}(U) = (U n_i)\sigma(U s_{i,j}) = \sigma_I (n_i \cdot U'e)(s_{i,j} \cdot U'e) = \sigma_I \cos(\phi_i(U)) \cos(\psi_{i,j}(U)) \tag{4}$$

Here $\phi_i(U)$ and $\psi_{i,j}(U)$ stand for the orientation dependent angle between the stress direction $e$ and the slip plane normal $U n_i$ and the slip system direction $U s_{ij}$, respectively. The maximal shear stress $\tau(U)$ on one of the slip systems then again is obtained by maximizing the modulus of $\tau_{ij}(U)$ and $m(U) = \tau(U)/\sigma_I$ is the orientation dependent Schmid factor. Note that the middle equation in (4) indicates that we can equivalently transform the stress direction e with $U'$ and leave lip plane normals and slip system directions unchanged.

At this point we recall that orientations $U$ and Schmid factors $m(U)$ have been determined experimentally for the surface grains from which fatigue cracks initiated by EBSD. Figure 5 shows that in the case of the experimental data, the scatter in LCF life is reduced when passing from the amplitude stress $\sigma_{Ia}$ to the related maximal traction amplitude stress $\tau_a(U) = m(U)\sigma_{Ia}$.

As stress-life curves often are not the preferred option in gas turbine design, we now introduce a model that allows to calculate Schmid factor adjusted strain amplitudes. To this aim let

$$\epsilon_a = R(\sigma_a) = \frac{\sigma_a}{E} + \left(\frac{\sigma_a}{K}\right)^{1/n'} \tag{5}$$

be the Ramberg Osgood relation and let $\sigma_a = R^{-1}(\epsilon_a)$ be the inverse. Given the Schmid factor m, we then define the Schmid factor adjusted strain amplitude as

$$\epsilon_a(m) = R\left(\frac{m}{\lambda} \cdot R^{-1}\epsilon_a\right) \tag{6}$$

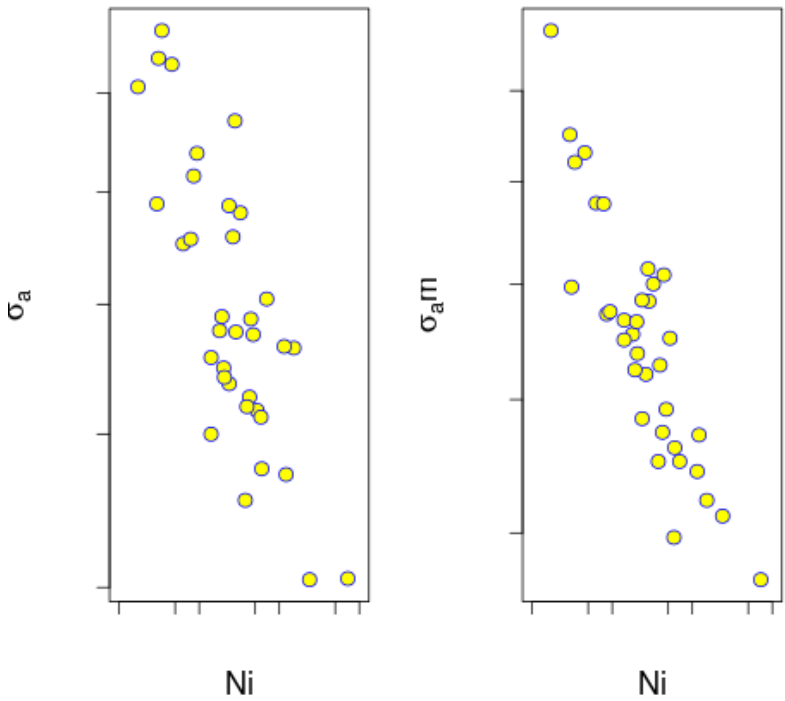

Fig. 5: Amplitude stress (left) and maximal traction amplitude stress vs cycles to crack initiation $Ni$ in double logarithmic printing. $Ni$ scaling on both panels is identical.

where $\lambda = 0.4524$ is some numerical constant obtained as the expected value of Schmid factors under isotropic random orientations, see the following section.

Again a remarkable reduction in the scatter in Ni is observed, in particular for high strain amplitudes. A life prediction model in the Coffin-Manson Basquin style can now be fitted to this data in the usual style, namely

$$\epsilon_a(m) = CMB(N_i) = \left(\frac{\sigma_f}{E}\right)(2N_i)^{-b} + \epsilon_f'(2N_i)^{-c} \qquad (7)$$

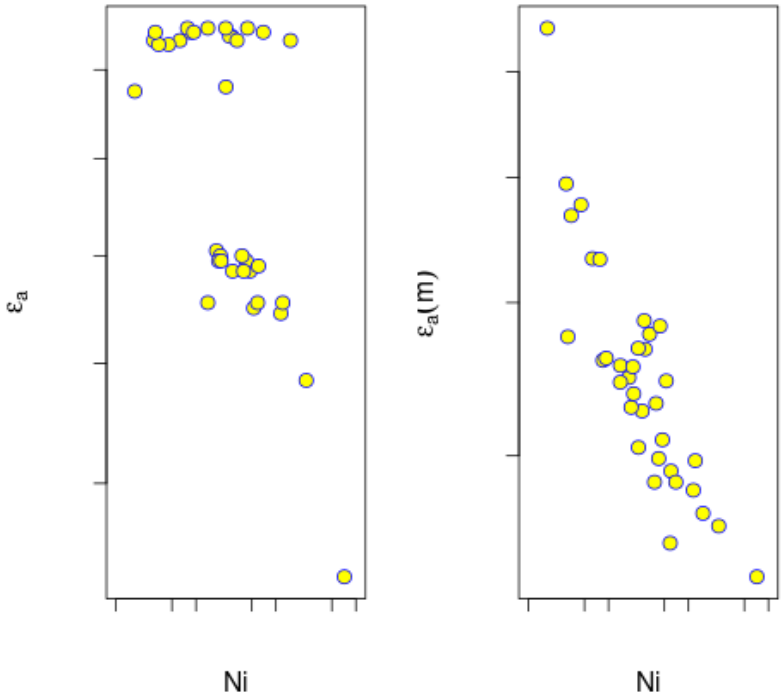

Fig. 6: Amplitude strain (left) and Schmid factor adjusted amplitude strain vs cycles to crack initiation $Ni$ in double logarithmic printing. $Ni$ scaling on both Panels is identical.

Hence the inverted relation $N_i(m) = CMB^{-1}(\epsilon_a(m))$ gives a predicted life that depends on the Schmid factor $m$. Figure 7 shows the fit to the experimental data.

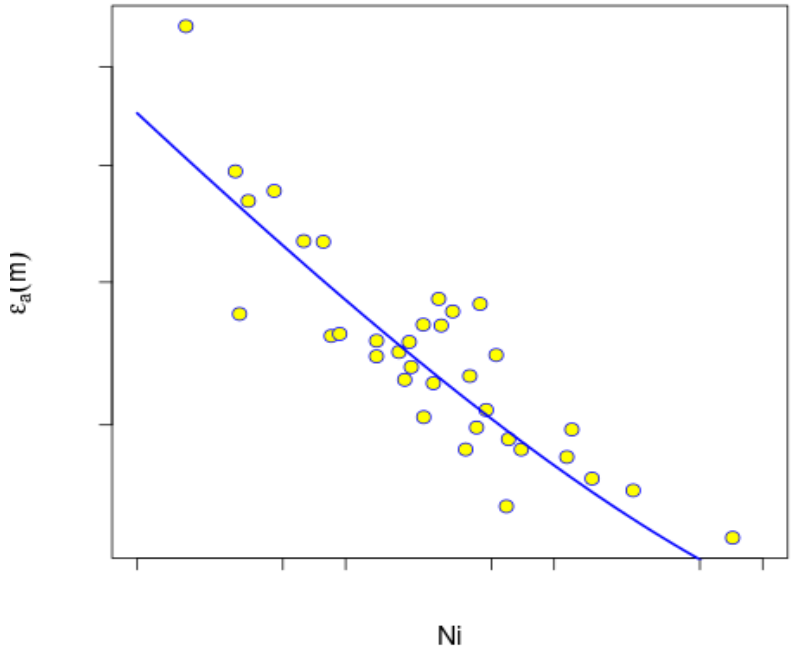

Fig. 7: Coffin-Manson Basquin relation (7) fitted to the experimental data for the Schmid factor adjusted strain amplitudes. Both axes are logarithmic.

**PROBABILISTIC SCHMID FACTORS AND SCATTER IN LCF LIFE**

In order to determine the probability distribution of $N_i(m(U))$ under isotropic grain orientation, we need to fix the isotropic probability law for $U$. In mathematics, this is known as the Haar measure on the orthogonal group [15]. In the present case where only uniaxial stress states $\sigma = \sigma_l e \otimes e$ are considered, the resulting probability law for the Schmid factors $m(U)$ only depends on the random direction $e(U) = U'e$, cf. Equation (4). Under the Haar measure, this random direction is uniformly distributed over the sphere embedded in three dimensional space. We therefore proceed with the following algorithm that is repeated until a sufficiently large Monte Carlo sample size has been realized:

1. Generate a random direction $e$ using pseudo random number generator;
2. For each slip plane normal and slip direction in crystallographic standard orientation calculate scalar products $e \cdot n_i$ and $e \cdot s_{ij}$ to obtain $m$;
3. Use Equations (6) and the inverse of (7) to calculate $Ni(m)$.

For the generation of random directions *e* that are uniformly distributed over the sphere we use the rotation invariance of the three dimensional multivariate normal distribution $N(0,1_{3x3})$ as follows

1. Generate independent pseudo random numbers $X_1$, $X_2$ and $X_3$ with standard normal distribution;
2. Calculate the random direction *e* by $e_i = \frac{X_i}{(X_1^2+X_2^2+X_3^2)^{\frac{1}{2}}}$, $i = 1,2,3$.

The resulting density for the distribution of Schmid factors is given in Figure 8.

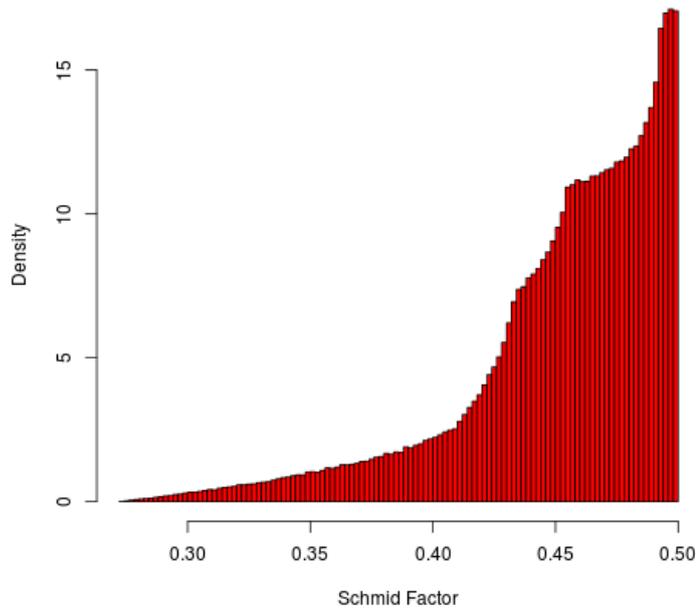

Fig. 8: Monte Carlo simulation of the probability density of Schmid factors under isotropic distribution of orientations using $10^6$ Mote Carlo samples.

Using this Monte Carlo sample, the Monte Carlo scatter of LCF life at a given average strain level can now be generated. Results are displayed in Figure 9. Note that in this simulation we exclusively consider the effect of random Schmid factors and neglect the residual scatter that is visible in the right panel of Figure 7.

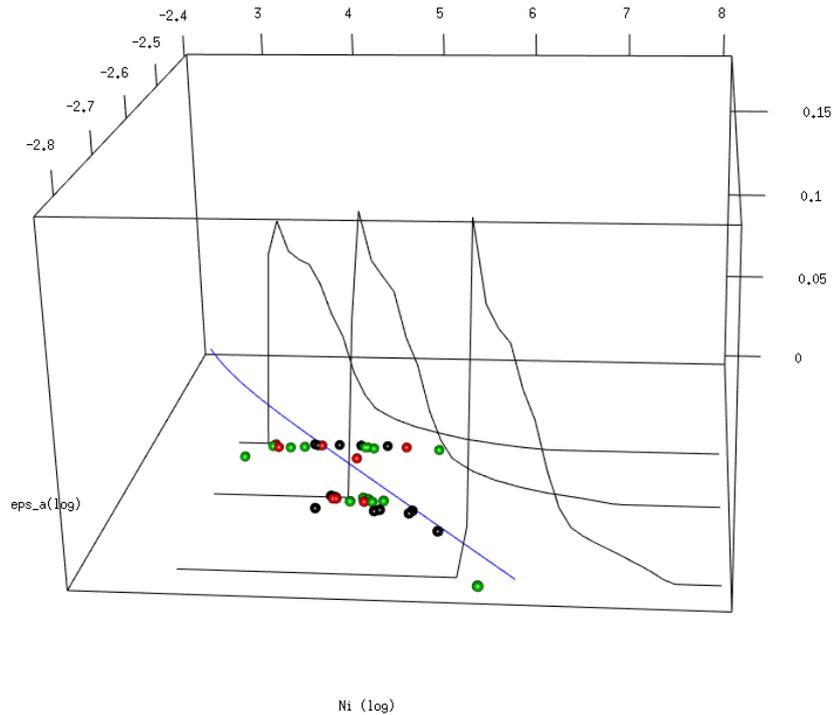

Fig. 9: LCF life distribution as predicted from the isotropic Schmid factor distribution for three average strain ampitude levels. The blue line marks the average strain to life CMB relation. Points display measured strain amplitude (non modified) vs $Ni$ values. Different specimen geometries are marked by different point colors: G1 green, G2 red, G3 black.

Interestingly, the range of scatter in LCF life due to Schmid factors has about the order of magnitude observed in the experiment. Note that single points outside of the scatter band are not problematic as we neglected the residual scatter visible in Figure 7 in this modelling step.

**EXTREME VALUE STATISTICS AND SCHMID FACTORS**

In this section we present some evidence that is against a naïve usage of the Schmid factor distribution as a LCF life distribution 'grain-by-grain'. In order to approach this problem, we first look at the grain surface distribution for the cast plates, see Figure 10.

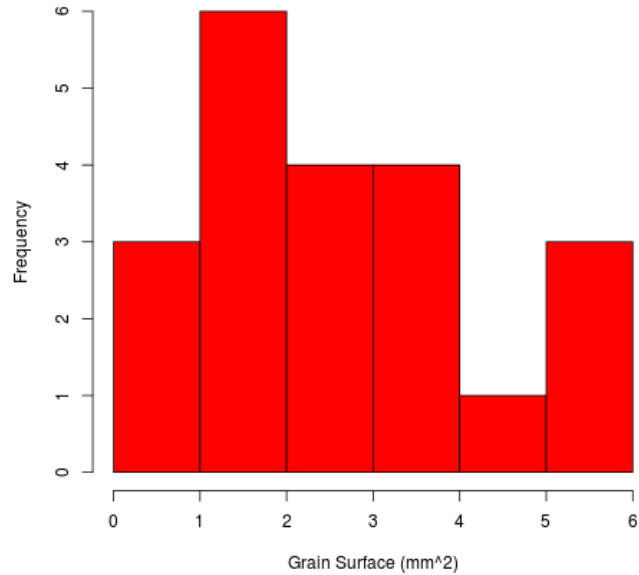

Fig. 10:  The grain surface distribution for the specimen obtained from microsections.

From the above data we conclude that an average of 138 grains will be needed to cover the gauge surface of geometry G1, 17 for G2 and 239 for G3. Hence, if we suppose that crack initiation occurs at the grain with the highest Schmid factor and if we assume independence of grain orientations, we have to replace the Schmid factor distribution in Figure 8 with the associated maximum value distribution for 138, 17 and 138 independent realizations, respectively [16].

Such maximum distributions however have a strongly reduced scatter, confer Figure 10 for the case of the smallest Geometry G2. This in turn leads to an enormously reduced scatter in LCF life well below an order of magnitude compared to the experimentally observed one. It is therefore worth to carefully check the hypothesis of this weakest link approach.

In particular, if it was true that the component fails at the grain with the highest Schmid factor, grain orientations at grains that did actually develop a fatigue crack should show a strong shift towards high Schmid factors when compared with the distribution obtained from isotropic random grain orientation. Figure 12 however shows no such behavior in the empirical data, where even a shift to lower Schmid factors becomes visible. Statistical testing of this effect with Kolmogorov-Smirnov test statistics reveals D=0.27 and a p-value of 1.2%. Note that this test is performed with respect to the ordinary Schmid factor distribution. Tests with the maximum distribution will be much more drastic. In conclusion the experimental evidence is clear enough to reject the hypothesis that LCF crack initiation in Specimen is causally connected to a high Schmid factor. This does not stand against the previous observation that Schmid factors do have an effect on the number of cycles to crack initiation. Simply, a bad grain orientation alone seems to be insufficient to initiate a crack.

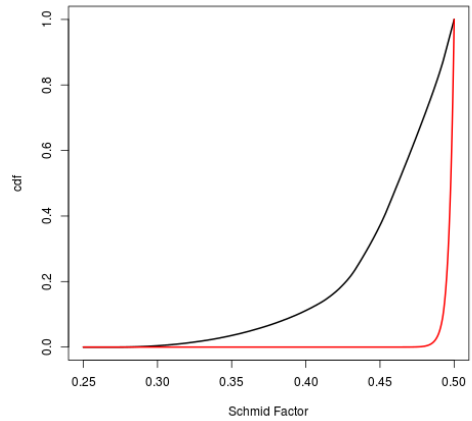

Fig. 11: Schmid factor probability distribution function and probability distribution function of maximum independent Schmid factors with 17 grains (G2).

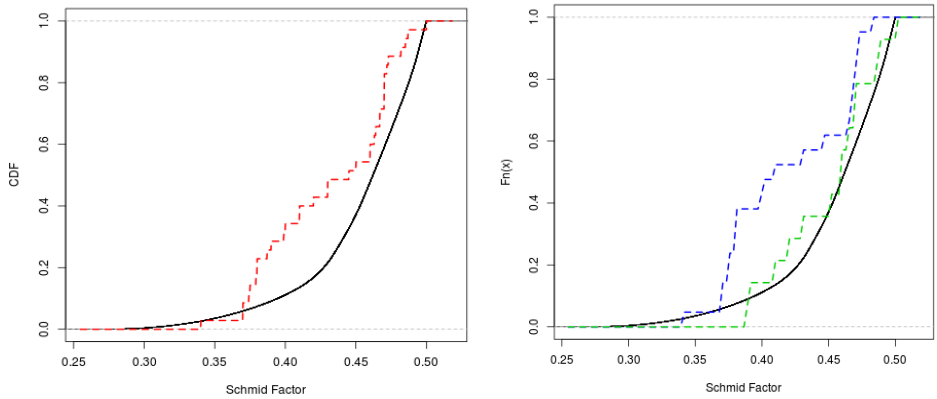

Fig. 12: Empirical cumulative distribution function (ecdf) of Schmid factors of grains that did develop a LCF crack (red) compared with the cumulative probability distribution function of Schmid factors under random grain orientation (left). Ecdf for amplitude strains with $\epsilon_a$ smaller than 0.2% (green) and larger than 0.2% (blue) as compared with the ecdf under random grain orientation (right).

A differentiation of the Schmid factor distribution with respect to amplitude strain levels however shows that at lower amplitude strains Schmid factors become larger and hence more important. Consistently with the observed complete plastification at high strain levels in Figure 3, it seems that also slip systems with non maximal Schmid factor can contribute to crack initiation explaining the left shifted empirical cumulative distribution function at high strain levels in Figure 12 b). At moderate amplitude strain, the empirical cumulative distribution function coincides remarkably well with the one of the uniform grain orientation.

**SUMMARY AND CONCLUSIONS**

In this work we investigated the effect of grain orientation to LCF life for the Ni base superalloy RENE 80 with fcc lattice structure in the γ-matrix. While we have seen that grain orientation measured by EBSD at crack initiation location is related to the LCF life via Schmid factors, considerations from extremal value statistics show that a 'bad' Schmid factor alone is insufficient to initiate a LCF crack.

It has been however been observed that the empirical cumulative distribution function of Schmid factors of grains from which LCF cracks originate is right shifted towards that of maximal Schmid factors with random grain orientation when the amplitude strain goes down. One may speculate that the process of right shifting of Schmid factors at grains where LCF cracks initate will go on and thus get closer to the extremal value distribution in Figure 11 when considering even lower amplitude strains. The importance of maximal Schmid factors for crack initiation thus might grow with decreasing stress. This hypothesis however goes beyond what can be proven based on our present experimental data.

It also seems to be necessary to further investigate effects in the random grain structure of Ni-based superalloys that are more directly related to crack initiation [17]. Natural candidates would be relative orientation of activated slip systems to the surface, grain size and the configuration of the surrounding grain.

**Acknowledgements**

This work has been supported by the German federal ministry of economic affairs (BMWi) and Siemens Energy via an AG Turbo grant.

**Corresponding author:** hanno.gottschalk@uni-wuppertal.de